\documentclass[reprint, pra, floatfix, superscriptaddress, amsmath, amssymb, aps]{revtex4-2}
\usepackage{siunitx}
\usepackage[version=4]{mhchem}
\usepackage[T1]{fontenc}
\usepackage{graphicx}
\usepackage{dcolumn}
\usepackage{bm}
\begin{document}


\title{InAs/InP quantum dot based C-Band all-fiber plug-and-play triggered single-photon source integrated using micro-transfer printing}

\author{Marek G. Mikulicz}
\email{marek.mikulicz@pwr.edu.pl}
 \affiliation{Department of Experimental Physics, Faculty of Fundamental Problems of Technology, Wrocław University of Science and Technology, Wybrzeże Wyspiańskiego 27, 50-370 Wrocław, Poland}
 
\author{Paweł Mrowiński}%
 \email{pawel.mrowinski@pwr.edu.pl}
\affiliation{Department of Experimental Physics, Faculty of Fundamental Problems of Technology, Wrocław University of Science and Technology, Wybrzeże Wyspiańskiego 27, 50-370 Wrocław, Poland}

\author{Paweł Holewa}
\affiliation
{Department of Experimental Physics, Faculty of Fundamental Problems of Technology, Wrocław University of Science and Technology, Wybrzeże Wyspiańskiego 27, 50-370 Wrocław, Poland}
\affiliation
{DTU Electro, Technical University of Denmark, 2800 Kongens Lyngby, Denmark}
\affiliation
{NanoPhoton-Center for Nanophotonics, Technical University of Denmark, 2800 Kongens Lyngby, Denmark}

\author{Kresten Yvind}
\affiliation
{DTU Electro, Technical University of Denmark, 2800 Kongens Lyngby, Denmark}
\affiliation
{NanoPhoton-Center for Nanophotonics, Technical University of Denmark, 2800 Kongens Lyngby, Denmark}
\author{Marcin Syperek}
\affiliation
{Department of Experimental Physics, Faculty of Fundamental Problems of Technology, Wrocław University of Science and Technology, Wybrzeże Wyspiańskiego 27, 50-370 Wrocław, Poland}
\author{Elizaveta Semenova}
\affiliation
{DTU Electro, Technical University of Denmark, 2800 Kongens Lyngby, Denmark}
\affiliation
{NanoPhoton-Center for Nanophotonics, Technical University of Denmark, 2800 Kongens Lyngby, Denmark}

\date{\today}

\begin{abstract}
Fiber-based long-haul quantum communication would greatly benefit from a robust and deterministically integrated source of quantum state. Here, we report the design, fabrication, and optical characterization of InAs/InP quantum dots in the InP H1 point defect 2D photonic crystal cavity, integrated with the standard single-mode fiber using a micro-transfer printing technique. The device was placed in a compact cryocooler maintaining a cryogenic temperature of 15 K with single-photon emission characterized by \(g^{(2)}(0)=0.14(14)\) and reliable and stable emission (intensity fluctuations given by a standard deviation $\sigma = 0.13$), so that an all-fiber based connection between two laboratory nodes through an open area was established and utilized for testing the quantum channel. In this way, we demonstrate a plug-and-play all-fiber single-photon source operating in the third telecom window, where standard telecommunication fiber networks can be used as a low-loss medium.

\end{abstract}
\maketitle


\section{Introduction}

Progress in solid-state system fabrication methods has significantly improved the quality of single-photon sources based on self-assembled semiconductor quantum dots (QDs)~\cite{Senellart2017}. These sources have reached very high levels of performance in quantum technology~\cite{Gschrey2015,Claudon2010,Somaschi2016}. The high-throughput processing of photonic micro- and nanostructures allowing integration of single QD emitters in a modified photonic environment~\cite{Lodahl2004} can enhance radiative recombination rates and emission directionality (contributing to overall source brightness~\cite{Tomm2021}), single-photon emission purity~\cite{Schweickert2018-px,Michler2002-do, Michler2000-xt, Schweickert2018,Miyazawa2016}, indistinguishability of consecutively emitted photons~\cite{Somaschi2016,Santori2002-uk,Hong1987-zu,Ding2016-xn, Wei2014,Thoma2016}, or the strength of the light-matter coupling~\cite{Lodahl2013} with the possibility of creating entangled quantum states of light~\cite{Ekert1991,Briegel1998,Gisin2007,Simon2007,Mower2013, Benson2000,Moreau2001,Stevenson2006}. Nanostructures exploiting the increase of local density of states are typically realized by fabricating mesas~\cite{Fischbach2017,Bremer2022,Schneider2018}, microlenses~\cite{Gschrey2015,Sartison2021}, micro-discs, 3D-printed micro-objectives~\cite{Fischbach2017}, photonic wires~\cite{Claudon2010,Bounouar2012,Munsch2013,Jns2017,Bulgarini2014,Holmes2014} micropillar cavities~\cite{Wang2019,Thomas2021,Somaschi2016}, photonic crystals~\cite{Lodahl2004,Pernice2012,Schrder2017}, circular Bragg gratings~\cite{Ates2012,Liu2019,Wang2019,Barbiero2022,Holewa2024}, tuneable cavities~\cite{Najer2019,Tomm2021} and hybrid structures~\cite{Northeast2021,Abudayyeh2021}. Alternatively, to improve the strength of the light-matter coupling, i.e. the ratio between the quality (Q) factor and the volume of the cavity mode, a cavity with extreme dielectric confinement can be used~\cite{Wang2018}. This modified photonic environment allows the realization of quantum communication~\cite{Vajner2022,Flamini2018,Northup2014,Gisin2007}, quantum teleportation~\cite{BassoBasset2021}, and as input of quantum states for photonic quantum computing~\cite{Ladd2010,Knill2001}.
An advantage of QDs is that they can be synthesized to emit light tuned to telecommunications windows, and connecting them directly to optical fibers is especially attractive because it eliminates bulky coupling setups, allowing the fabrication of compact devices~\cite{Schlehahn2015, Schlehahn2018,Musia2020,Gao2022}. Furthermore, the microscope objectives can collect luminescence signals within a specific field of view and numerical aperture, which limits the signal collection efficiency and requires active stabilization for long-lasting experiments or practical use. In contrast, direct fiber coupling, in principle, offers maximized collection efficiency and solid contact with the operating QD. Moreover, access to numerous emitters on a single chip via a directly coupled fiber array offers high scalability. For all-fiber configurations, the most straightforward approach is to integrate the chip with a QD directly with an optical fiber~\cite{Xu2007}. To increase coupling efficiency, the emitter (or cavity with an emitter) can be integrated with a standard, high-NA or lensed fiber, GRIN-lenses assembly~\cite{Northeast2021}, and glued to the microobjective~\cite{Bremer2020}. For more advanced structures, this approach can be improved with the pre-selection of the most suitable emitter before integration. The emitter can be deterministically integrated with a fiber, for example, with a focused ion beam (FIB)~\cite{Lee2019}, a modified 2D circular Bragg grating can be transferred with micro-transfer printing (µTP)~\cite{Jeon2022}, or using interferometric positioning over the mesa structure to attach the fiber core, which, however, limits the yield to only one structure per chip~\cite{Musia2020,Gao2022}. With the use of the µTP technique, it is in principle possible to integrate all structures from the source wafer, allowing for a high fabrication yield (high integration densities and efficient material utilization~\cite{Zhang2019}) and selectivity of pre-characterized devices; this also allows for more versatile integrated platforms.

The quantum communication protocols can operate at the telecom windows, multiple approaches are directed towards the 900-nm wavelength range~\cite{Bremer2020, Northeast2021} and in the second telecom window centered~\cite{Lee2019,Jeon2022,Musia2020,Gao2022}, where the GaSb/AlGaSb QDs are attractive for long spin coherence times~\cite{Michl2023}, however, all-fiber demonstrations at 1550 nm are still missing. In this context, InAs/InP quantum dots are especially attractive, as they do not require substantial modifications to the growth conditions, known for the GaAs-based system, such as strain control~\cite{Guffarth2001, Tatebayashi2001} or metamorphic buffer layer~\cite{Semenova2008,Paul2017,Portalupi2019}, to fit the 1550 nm.~\cite{berdnikov2023fine} InAs/InP QDs are characterized by the high purity of single-photon emission~\cite{Miyazawa2016}, indistinguishability~\cite{Holewa2024,Rahaman2024}, and by the emission of polarization-entangled photon pairs~\cite{Kors2018}. For the third telecom window, advanced metropolitan quantum networks have already been established~\cite{Gyger2022} with InAs/GaAs quantum dots, and InP-based sources of ultrahigh single-photon purity are also used~\cite{Takemoto2015}, although they are not directly fiber-coupled. In this telecom window, InAs QD can also be interfaced with an atomic quantum memory~\cite{Thomas2024} for scalable quantum networks. Direct integration with an optical fiber allows to use cost-effective and compact cryocoolers that are available on the market, and in spite of inevitable vibrations inside that hinder the use of free-space optics, the use of QD sources directly coupled to optical fibers is needed.

Here, we report on the design, fabrication, and optical investigation of an all-fiber coupled single-photon source operating at the telcom C-band based on InAs/InP quantum dots integrated with an H1 point defect photonic crystal (PhC) cavity. The device was integrated using µTP with a standard single-mode optical fiber and placed in a compact cryocooler at 15 K, showing triggered single-photon operation.

\section{Device Design and Fabrication}
\subsection{Device Design}
\begin{figure}[ht]
\includegraphics[width=0.5\textwidth]{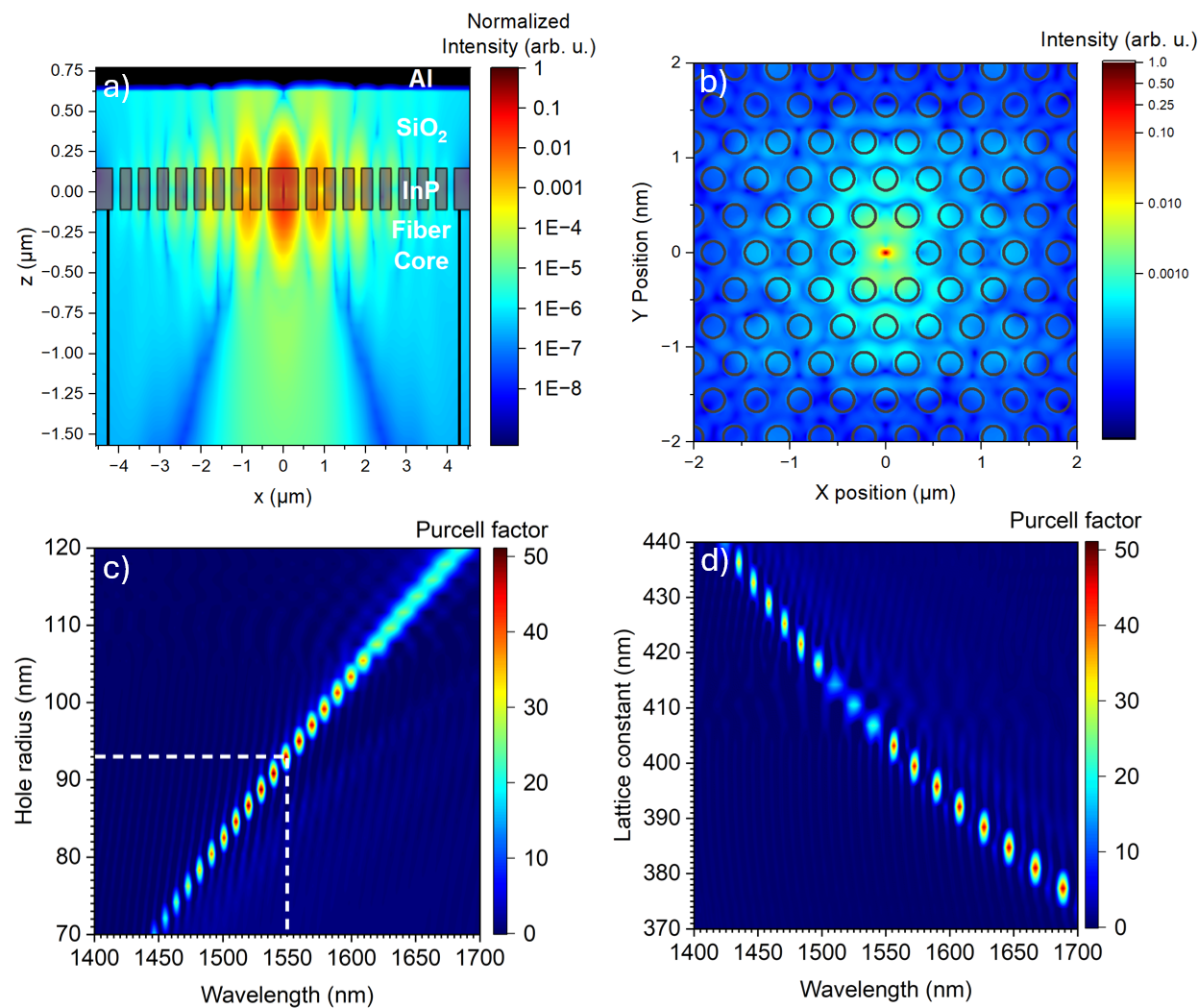}
\caption{FDTD simulations a)-b) cross-sectional electric field intensity distribution in logarithmic scale showing a) the directional emission to the fiber core in z direction b) light interaction in-plane of the H1 PhC cavity. c) Dependence of the Purcell factor on the number of hole radius (calculated for a~=~420~nm) and d) lattice period (calculated for  r~=~80~nm).}
\label{fig:Simulations}
\end{figure}

To increase QD extraction efficiency and investigate cavity quantum electrodynamics with a QD, the source is typically placed in a resonator that can be realized in the form of an H- or L-type 2D photonic crystal cavity~\cite{Saldutti2021,Phillips2024}. Fabrication of the PhC cavity shown in Fig.~\ref{fig:Fig_fabrication} was preceded by numerical simulations of the H1 PhC cavity using finite difference time domain (FDTD) numerical simulations with a finite element solver~\cite{Lumerical}.  This enables the numerical examination of light transmission through the integrated structure by resolving Maxwell's equations in the time domain, where the focus was placed on finding the fundamental cavity mode centered at 1.55 µm wavelength with a Purcell factor up to 50. Simulation of the QD emission was realized by placing the in-plane dipole source at the center of the cavity structure. The boundary condition surrounding the simulation area is defined as a perfectly matched layer~\cite{Brenger2007}. The thickness of the layers was fixed at values described in the chip fabrication section, which matched the fabricated sample. The vertical cross-section of the cavity and the intensity profile of the electric field in the simulation model are shown in Fig.~\ref{fig:Simulations} a). The field is mostly localized in the center of the cavity and reflected from the Al mirror toward the core of the fiber. The fundamental mode at 1.55 µm with a Purcell factor of about 50 was found for the specific cavity with a period = 420 nm and a hole radius = 93 nm, which is highlighted in Fig.~\ref{fig:Simulations} c) with the white dashed lines. Figs.~\ref{fig:Simulations} c) and d) show the corresponding calculated resonances as a function of period and hole size (hole radius). The calculated transmission of the dipole field in resonance with the cavity mode to the fiber core fundamental mode is around \SI{27}{\percent}. As seen for scanning parameter space, we observe, besides the fundamental cavity mode, also higher cavity modes with a weaker Purcell factor, to which the dense QD emission lines can also couple.

\subsection{Chip fabrication}
\begin{figure}[ht]
\includegraphics[width=0.5\textwidth]{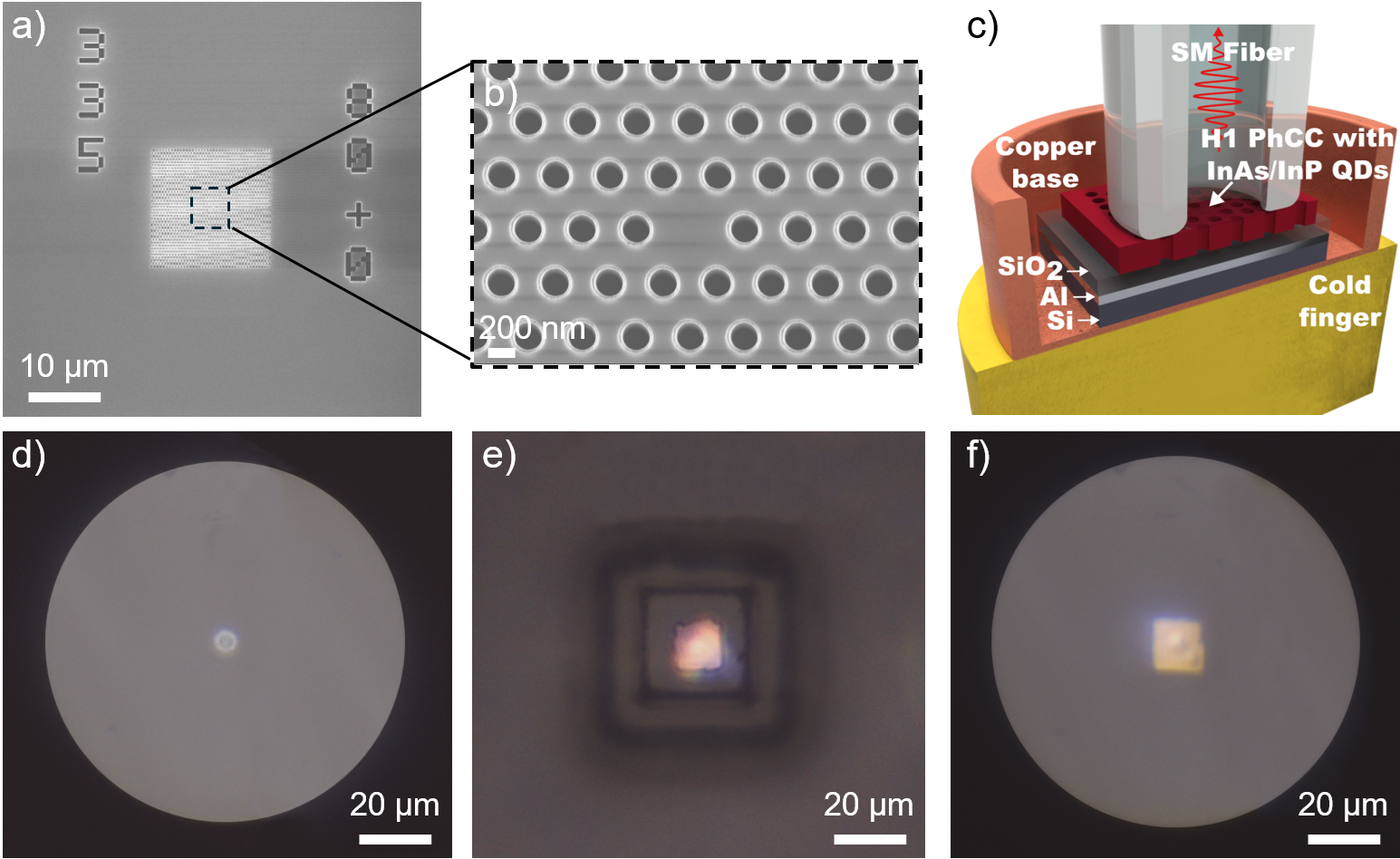}
\caption{Integration of the single 2D photonic microstructure into a cleaved SM fiber. a)-b) SEM images of a fabricated H1 2D PhC cavity prior to membranization. c) Conceptual visualization of the cross-section of the fabricated device d) View of the fiber core before transfer through micro-transfer setup without PDMS stamp. e) View of the structure on the PDMS stamp during transfer. f) Image of the structure transferred to the fiber core.}
\label{fig:Fig_fabrication}
\end{figure}

In our device design, the emitter is an InAs/InP quantum dot grown by metal-organic vapor-phase epitaxy (MOVPE)~\cite{holewa2020prb,berdnikov2023fine}. During growth on a (001)-oriented InP substrate at \SI{610}{\degreeCelsius} there is deposited a 500 nm InP buffer layer, a 200 nm-thick In\textsubscript{0.53}Ga\textsubscript{0.47}As sacrificial layer lattice-matched to InP, and a 244 nm thick InP layer, followed by 0.93 monolayers of InAs for nucleation of QDs in the sub-critical regime Stranski-Krastanov growth mode achieving high density. By tuning the V/III ratio, the density of QDs can be controlled. The deposition of a 244 nm-thick InP cap layer finishes the growth. 500 nm of \ce{SiO2} is a release layer deposited by PECVD, and with an e-beam evaporator, 120 nm of Al is deposited.  These two layers serve a dual function: first, after the etching process, it allows separation of the device from the source wafer; second,  \ce{SiO2} transparent at 1.55 µm enables the prospective 2D imaging~\cite{Holewa2024}. The wafer is spin-coated with benzocyclobutene (BCB) and then bonded to the Si wafer with high force (\SI{2}{kN}), at a temperature of \SI{250}{\degreeCelsius} under vacuum conditions in the wafer bonder. After bonding, the InP substrate and In\textsubscript{0.53}Ga\textsubscript{0.47}As stop layer are removed using \ce{HCl} and \ce{(10\%)H2SO4$:$H2O2}, respectively~\cite{Holewa2022}. The 96-nm thick SiN layer is deposited on the InP membrane. This layer acts not only as as a mask for electron-beam lithography but also as a stress management layer by introducing tensile stress. Consecutively, a series of designed H1 point defect 2D PhC cavities is patterned with electron-beam lithography. The holes' radius was set to 80 nm, 100 nm, or 120 nm, and the lattice constant varies in the range from 335 to 435 nm with a step of 10 nm. After dry etching of InP with \ce{HBr} in the inductively coupled plasma reactive ion etching, the structure was membranized in a BHF wet etch by removing the release layer \ce{SiO2} underneath a coupon. SEMs of the exemplary PhC cavity taken before membranization are shown in Fig.~\ref{fig:Fig_fabrication}. After membranization, the cavities were characterized at room temperature in µPL setup, with the specific results for a device with a hole radius of 80 nm and a lattice constant of 405 nm described in Section IIIB.

\subsection{Micro-transfer printing}
In recent years, a highly deterministic approach in heterogeneous integration was demonstrated via the pick-and-place technique, where the individual structure is picked from a source and placed on a target structure~\cite{Yoon2015, Zhang2019,Roelkens2023}. A potential method for the integration of III-V material to fiber is µTP, which involves the pickup and transfer of mm to µm-sized elements, from a source wafer to a target wafer with high alignment accuracy using a polydimethylsiloxane (PDMS) stamp~\cite{Menard2004,Meitl2005}. The thickness of the elements may vary from a sub-micrometer to 20 µm. This technique offers benefits, such as high integration densities and efficient material utilization~\cite{Zhang2019}. However, careful material selection and processing are essential due to the risk of mechanical stress-induced damage to fragile released epitaxial layers during the transfer process, which can result in device degradation and collapse of the coupons. Typically, a release interlayer is used, such as \ce{InGaAs}, \ce{InAlAs} or \ce{SiO2}~\cite{OCallaghan2017, Zhang2019}. Spincoating BCB on the target wafer is commonly used to increase surface flatness and as an adhesive for bonding~\cite{Niklaus2006}. Alternatively, if the released surface exhibits sufficient flatness and smoothness (typically with the root mean square average of height deviation taken from the mean image data plane, Rq < 1 nm), the devices can be integrated directly onto the target substrate by van der Waals adhesion forces~\cite{Loi2016}. This adhesive-free approach facilitates potential electrical connections and increases heat dissipation for the devices. In AFM, we investigated the bottom of a non-perforated part of our structure (not shown here), and the Rq has a value between 1.0 and 1.8 nm. In the case of our design, our transfer yield is close to 100\% with 20 out of 21 successfully transferred devices. Additionally, the use of BCB can decrease the Q factor of H1 2D PhC cavity by about a factor of two as a result of decreasing the indices contrast~\cite{Kicken2008}. Micro-transfer printing was done using a commercially available X-Celeprint machine. Preselected 2D PhC cavities were picked up with a 50x50 µm PDMS stamp and placed on a fiber core, which was cleaved and cleaned before transfer with isopropanol. The PhC cavity structure is placed directly on the standard Corning\textsuperscript{\textregistered} SMF-28 fiber (core diameter \SI{8.2}{\micro m}) without any adhesive layer. The mode wavelength of the selected cavities at room temperature was around 1.57-1.58 µm, in order to compensate for low-temperature blueshift. The fiber cable can be illuminated from the other port to increase the contrast between the core and the cladding and to facilitate the cavity alignment. The microscope images taken at different steps of this µTP process are shown in Figs.~\ref{fig:Fig_fabrication} d) - f). The high magnification and stability of our setup allow for precise placement of the structure on top of the fiber core with a precision greater than 200 nm, and in specially modified setup the µTP accuracy can be as low as 50 nm~\cite{Katsumi2018} or even allowing direct docking contact in two axes~\cite{Lu2023}, depending on the integration method. After micro-transfer printing, the fiber with PhC cavity was further secured to the Si chip using a temporary bonding adhesive (Crystalbond\texttrademark~555), which had deposited 0.5~µm \ce{SiO2} and 120-nm Al layers acting as a transparent spacer and backreflection layer (to control interference) with conceptual layer cross-section shown on Fig.~\ref{fig:Fig_fabrication} c). To introduce structural robustness for fiber manipulation and installation in a cryocooler, we subsequently secured the chip to the fiber with UV-cured epoxy and bonded the whole assembly to a copper base using aluminum epoxy. This additionally allows efficient heat transfer to the cryocooler cold finger and better temperature stability.

\section{Device characterization}
\subsection{Experimental setup}

\begin{figure}[ht]
\includegraphics[width=0.5\textwidth]{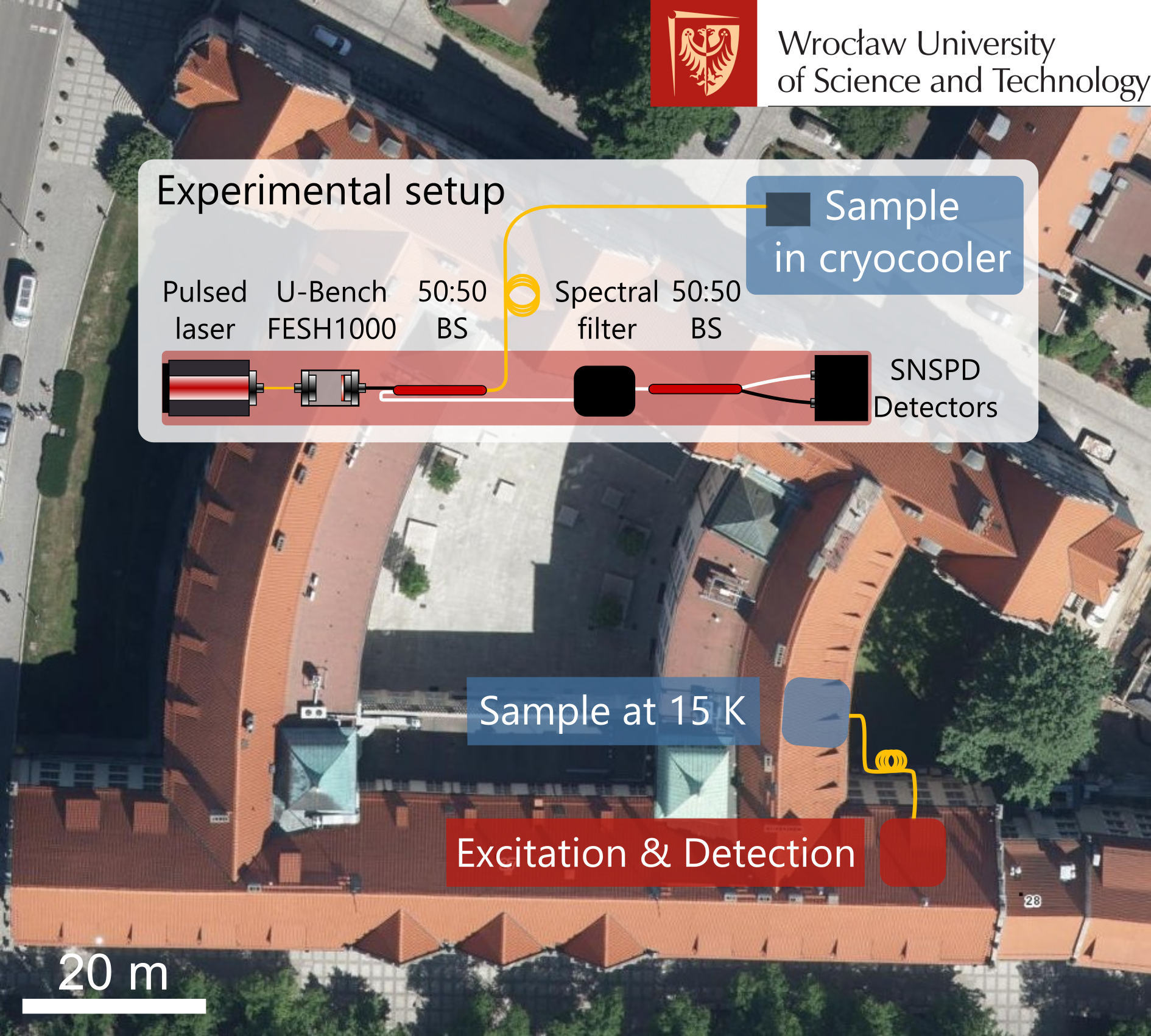}
\caption{All-fiber experimental setup with coupled single-photon system. In the photo background is the location of the two laboratory nodes used for the experiment. In the inset there is a schematic of the all-fiber path from the excitation laser to the source and to the detectors.}
\label{fig:PWr}
\end{figure}

For low-temperature microphotoluminescence spectroscopy, the sample was kept in a closed-cycle cryocooler at 15 K. The sample was excited with a fiber-coupled 805-nm pulsed laser (PicoQuant LDH-D-C-810) with a repetition rate of up to 80 MHz (50 ps-long pulses), and for continuous-wave experiments, the sample was excited with a 787-nm laser (Coherent CUBE) also fiber-coupled. For high excitation powers, spectral analysis was done on a 0.5 m focal length monochromator (Princeton Instruments) with an InGaAs multichannel array detector. For the measurements that employed the superconducting nanowire single-photon (SNSPD) detectors, it was necessary to filter the laser background emission in the longer wavelengths with a shortpass (1000 nm) filter mounted onto a fiber-to-fiber U-Bench. The fiber-coupled laser was connected to the 50:50 1x2 wide-band fiber coupler (Thorlabs TW1550R5F1) and connected to a 30 m stainless steel shielded SM fiber (SMF-28-J9) that was guided outside the building, connecting two laboratory nodes. In a different laboratory, the shielded fiber was connected to the cryocooler (CTI-Cryogenics, with dimensions of 30x40x25 cm) with the sample mounted inside through a fiber feed-through. The cryocooler was connected to the water-cooled compressor and evacuated with a turbo pump. The emission was collected from the other arm of the 50:50 beam splitter at the excitation and detection node, and the spectra were filtered with the polarization-insensitive electric tunable fiber optical filter of \SI{0.4}{nm} (Gaussian-shaped) FWHM bandwidth over the tuning range of 1520-1580 nm centered at 1550 nm and detected on the SNSPD detectors. The autocorrelation measurement was performed in the Hanbury Brown-Twiss configuration, where the signal after the fiber filter was split into two detection channels with an additional 1x2 50:50 fiber beam splitter. A schematic of the experimental setup is shown in the inset of Figure~\ref{fig:PWr}.

\subsection{Optical characterization}
\begin{figure}[ht]
\includegraphics[width=0.5\textwidth]{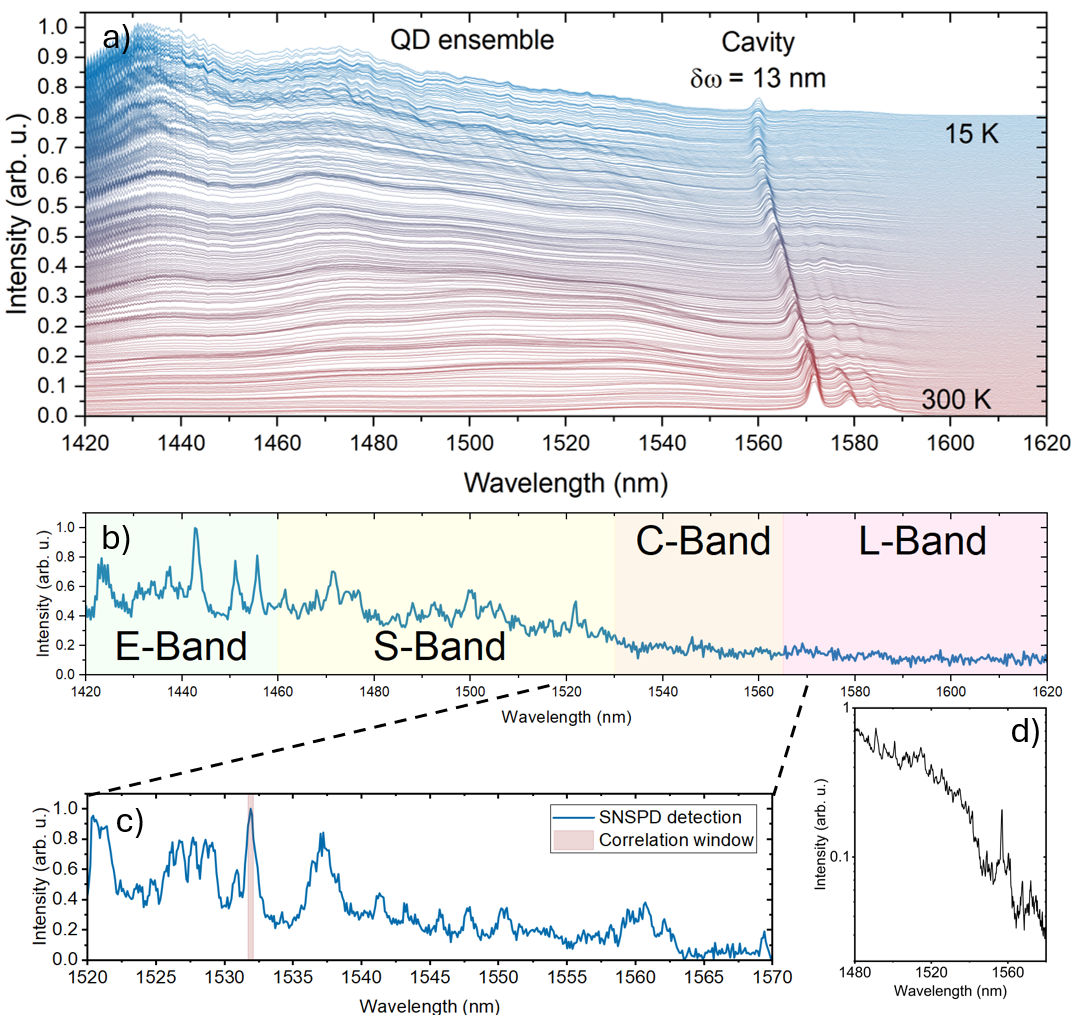}
\caption{Optical characterization of the device. a) Temperature-driven evolution of PL spectrum from 15 to 300 K for high power. b) Wide-range PL spectrum taken at low temperature and at low excitation power of a fiber-integrated PhC cavity. c) higher resolution spectra of the same device as in panel b), recorded using SNSPD, d) a different device with transition on cavity mode with linewidth more than two times narrower than that of the cavity mode.}
\label{fig:Temp_series}
\end{figure}

We first investigated the device using high excitation power. The temperature series is shown in Fig.~\ref{fig:Temp_series} a). At room temperature, we observed both the broad emission from the QD ensemble and the resonant cavity mode at 1573.4 nm. Although in this device we do not observe higher order cavity modes, for some devices they were identified in the spectra at shorter wavelengths. By cooling the sample down to 15 K, we observe a blueshift of the QD emission due to a Varshni-like shift of the bandgap and a blueshift of the cavity mode by 13 nm. The blueshift in the cavity mode is due to the temperature-dependent change in the refractive index~\cite{PhysRevApplied.21.044001} and thermal expansion of the cavity. Furthermore, we observe (at 15 K) an increase in the cavity Q factor (frequency-to-bandwidth ratio) from 620 to 750. At a low temperature of 15 K, the power was then decreased to observe single QD-like transitions in the telecom bands shown in Fig.~\ref{fig:Temp_series} b). To characterize the source in time-resolved experiments, the average laser power was set at 146 nW before the cryocooler fiber feedthrough, and the integrated spectrum was measured on the SNSPD detector as shown in~\ref{fig:Temp_series} c). Due to the diverse range of excited and ground states of high-density QDs, the number of excitonic transitions blends into a quasi-continuous spectrum, leading to the emission of quantum dot background radiation that undesirably feeds the fundamental mode of the cavity~\cite{Winger2009}. Figure~\ref{fig:Temp_series} d) shows the spectra of a different device with a Purcell-enhanced QD transition, which exhibits a linewidth more than twice as narrow as the cavity mode. This measurement was taken with a higher excitation power compared to that used in Fig.~\ref{fig:Temp_series}~c).

\begin{figure}[ht]
\includegraphics[width=0.5\textwidth]{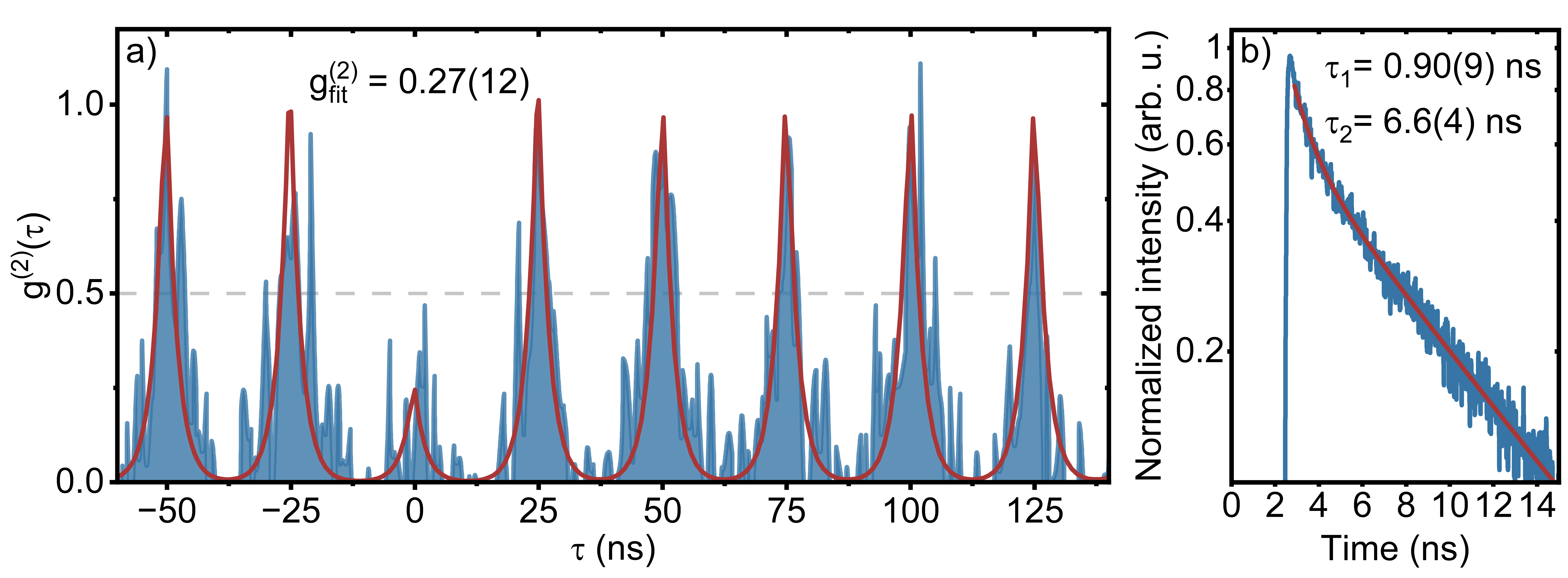}
\caption{Time-resolved experiments a) Second-order autocorrelation measurement with pulsed (triggered) optical excitation of 146 nW and 40 MHz repetition rate with background subtracted. The dip at zero delay time confirms that the emitter is a single photon emitter. The blue curve is the experimental data and red lines correspond to fitted curves showing $g_{fit}^{(2)}(0)=\SI{0.27(12)}{}.$ b) Measured transition decay profile of the quantum emitter within the same spectral window as for autocorrelation measurement. The decay curve is fitted using the double exponential function to estimate the decay constants of fast 0.90(9) ns and slow 6.6(4) ns.}
\label{fig:g2}
\end{figure}

The QD emission line at 1532 nm marked in Fig.~\ref{fig:Temp_series} c) was selected for the autocorrelation measurements. The high signal-to-noise ratio, the higher intensity than other lines, and the C-band operation make it an optimum choice. It also matches the parameters of the \ce{^{4}I_{13/2}\xrightarrow{\hspace*{0.2cm}}^{4}I_{15/2}} transition of erbium(III) ions~\cite{Wong2004}. For the line in Fig.~\ref{fig:Temp_series}~d) the uncorrelated background contributes to the multiphoton events and hinders the proof of its single-photon emission for this case; the low density of QDs is expected to overcome this constraint and allow pure single-photon Purcell enhanced emission from a single QD. For the line at 1532 nm to characterize the purity of single-photon emission, which is one of the crucial elements in the application of quantum computation and communication devices~\cite{Holmes2014,Santori2001,Brunel1999}, we used non-resonant pulsed excitation.  The photoluminescence decay measurement, which is shown in Fig.~\ref{fig:g2} b) reveals biexponential decay with fast (0.9-ns) and slow (6.6-ns) components. With typical lifetimes for InAs/InP QD ensembles of around 2 ns~\cite{holewa2020prb,Holewa2024,Phillips2024}, one possible explanation for a lifetime shorter than 1 ns is that the QD transition might be coupling to a higher-order cavity mode with a weak Purcell factor of around 2.

Due to the longer component of the photoluminescence decay, which may originate from the uncorrelated background emission~\cite{Winger2009}, we examined the HBT experiment with a laser repetition rate of 40 MHz to avoid the overlap of subsequent correlation peaks. The line was filtered with a tunable fiber-based filter, so the investigated window is within the spectral range marked by the pink area in Fig.~\ref{fig:Temp_series} c). The histogram data were fitted with a periodic exponential decay with the following formula~\cite{Miyazawa2016}:
\begin{equation}
    g_{fit}^{(2)}(\tau)=C_{bg}+g^{(2)}(0)e^{\frac{-|\tau|}{\tau_d}}+\alpha\sum_{n\neq 0}e^{\frac{-|\tau\pm nT|}{\tau_d}}
\end{equation}
From the normalized histogram of coincidences demonstrated in Fig.~\ref{fig:g2} a), we subtracted the coincidences caused by the background contribution $C_{bg}$, $\alpha$ is the amplitude of the pulses outside of the zero delay, and $\tau_d$ is the time constant describing decay with the value of \SI{2.2}{\nano s} from the histogram fit. The value of $g_{fit}^{(2)}(0)$ from the histogram fit is then $0.27(12)$. The measurement was carried out with a power close to the saturation level (power series not shown here), which allows for a faster observation of the minimum at $g^{(2)}(0)$ given the large background emission. In these experimental conditions, the fitted value of $g_{fit}^{(2)}(0)$ is rather high and with a power reduction, we expect a reduction of $g_{fit}^{(2)}(0)$~\cite{Musia2020}, but a decrease in power and consequently in the signal-to-noise ratio introduces difficulty in obtaining an accurate measurement. The $g_{fit}^{(2)}(0)$ does not reach the pure single emission value of $g_{fit}^{(2)}(0)=0$ and the background contribution to the single-photon purity can be accounted for using the formula:
\begin{equation}
    g^{(2)}_{corrected}(0)=\frac{C_N(\tau)-(1-\rho^2)}{\rho^2}
\end{equation}
with $\rho = (S)/(S+B)$ of 0.7 where $S$ and $B$ represent the signal and background count rates, respectively, of the observed line~\cite{Brouri:00}, and $C_{N}(\tau)$ is the measured number of coincidences normalized by the expected number of coincidences for a Poissonian source with the same intensity. This parameter enables the determination of the corrected value of \(g^{(2)}_{corrected}(0)=0.14(14)\) without background contribution.

\begin{figure}[ht]
\includegraphics[width=0.5\textwidth]{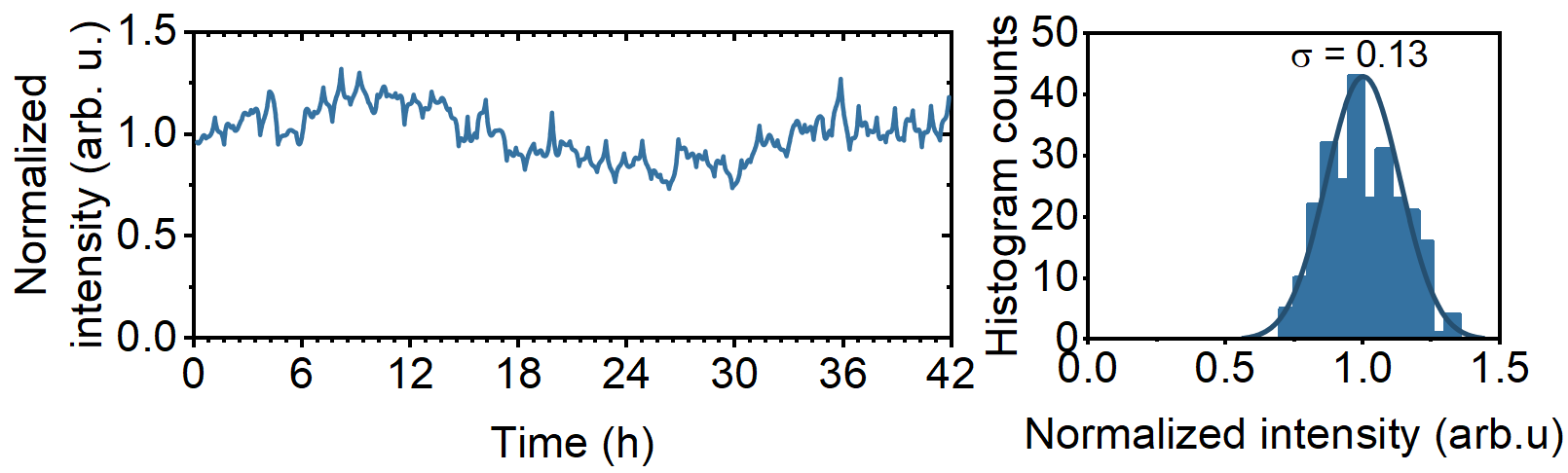}
\caption{a) Intensity plot of the normalized single-photon count rate of the device. b) Histogram of the normalized intensity showing the distribution of the signal intensity with a standard deviation value $\sigma = 0.13$.}
\label{fig:Fig_perf}
\end{figure}

The stability and operational readiness of a fiber-coupled QD source are important for real-world applications. We tested our device stability by monitoring the single-photon flux for more than 40 hours, where the normalized intensity plot of the single-photon count rate (the single-photon count rate) is shown in Fig.~\ref{fig:Fig_perf} a), showing long time-scale fluctuation, which may be caused by the thermal fluctuation of the cryocooler or the background fluctuations since the external fiber link is positioned in an open environment, rather than being concealed beneath the ground as typically for the fiber network. The intensity of the source remains the same after the thermal cycles of a cryocooler. The histogram of the signal distribution is shown in Fig.~\ref{fig:g2} panel b) and was fitted with a Gauss function with a standard deviation value $\sigma = 0.13$. The calculated signal fluctuation without long time-scale component is reduced to $\sigma = 0.09$.

\section{Results and discussion}
In this work, we have detailed the design, fabrication, and optical characterization of the triggered all-fiber single-photon source operating in the third telecom window based on high-density InAs/InP quantum dots. We have used an H1 point defect 2D PhC cavity to enhance photon coupling to the standard single-mode fiber, along with a metallic back reflector and a \ce{SiO2} layer spacer. The structures were integrated using the micro-transfer printing technique with a high yield. Operating between two distant laboratory nodes connected via commercially available single-mode fiber, the device maintains long-term stable emission and reliable performance with cooling cycles, demonstrating robust operation that is applicable to more real-world applications. The device shows single-photon emission with $g_{fit}^{(2)}(0)=\SI{0.27(12)}{}$ and without background contribution \(g^{(2)}_{corrected}(0)=0.14(14)\).

The used \ce{SiO2} release layer and metallic mirror at the bottom of the chip before transfer printing has the prospective for 2D imaging and deterministic localization of a single QD \cite{Holewa2024}. This would allow one to define a cavity with a pre-selected single QD in its center, which along using low-density QDs and would decrease the number of lines in the PL spectrum, thus decreasing the background contribution~\cite{PhysRevLett.103.207403} and improving the single-photon purity of the device. Reduction of background emission and improvement of single-photon purity can also be achieved by using the quasi-resonant excitation scheme~\cite{Gao2022}. In addition, the efficiency of light coupling can increase with improved surface finish (fiber end polishing)~\cite{Gharbia2004}. Compared to FIB integration, micro-transfer printing allows for introducing a feedback loop during the alignment of the structure where from the other side of the fiber the fundamental mode of the cavity could be observed in real-time, and micro-transfer printing can be adjusted to maximize the coupling. 

Our design preserves the ability of the device to generate single photons despite advanced processing, overcoming the challenges of precise alignment, mode mismatch, and mechanical stress when coupling to a fiber. This makes our integration framework suitable for applications beyond single-photon sources (used for quantum key distribution, quantum metrology, or photon-based simulations) that also require alignment with a fiber core or multiple cores and stable operation, such as lab-on-fiber sensing~\cite{Picelli2020, Vaiano2016}, highly sensitive measurement of refractive index and temperature~\cite{Jung2011, Park2014} and Raman scattering detection~\cite{Smythe2009}.

\begin{acknowledgments}
We acknowledge financial support from the National Science Centre (Poland) within Project No. 2020/39/D/ST5/02952, from the Danish National Research Foundation via the Research Centers of Excellence NanoPhoton (DNRF147), and from the European Union's Horizon Europe Research and Innovation Programme under the QPIC 1550 project (Grant Agreement No 101135785).
\end{acknowledgments}

\bibliography{Main_text}
\end{document}